\documentstyle[12pt]{article}
\begin{document}
\pagestyle{plain}
\setcounter{page}{1}

\begin{titlepage}

\begin{flushright}
PUPT-1726\\
NSF-ITP-97-113\\
hep-th/9710097
\end{flushright}
\vspace{10 mm}

\begin{center}
{\huge Brane-Waves and Strings}

\vspace{5mm}

\end{center}

\vspace{5 mm}

\begin{center}
{\large Sangmin Lee$^a$\footnote{sangmin@princeton.edu}, Amanda Peet$^b$\footnote{peet@itp.ucsb.edu}, and 
L\'arus Thorlacius$^a$\footnote{larus@feynman.princeton.edu}}

\vspace{3mm}

a) Joseph Henry Laboratories\\
Princeton University\\
Princeton, New Jersey 08544

\vspace{3mm}

b) Institute for Theoretical Physics\\
University of California\\
Santa Barbara, California 93106
\end{center}

\vspace{1cm}

\begin{center}
{\large Abstract}
\end{center}

\noindent
Recently, solutions of the Born-Infeld theory representing strings
emanating from a Dirichlet p-brane have been constructed.  We discuss the
embedding of these Born-Infeld solutions into the non-abelian theory
appropriate to multiple overlapping p-branes.  We also prove supersymmetry
of the solutions explicitly in the full nonlinear theory.  We then study
transverse fluctuations, both from the worldbrane point of view and by
analyzing a test-string in the supergravity background of a Dp-brane.  We
find agreement between the two approaches for the cases p=3,4.

\vspace{1cm}
\begin{flushleft}
October 1997
\end{flushleft}
\end{titlepage}
\newpage


\newcommand{\grad}{\nabla}
\newcommand{\tr}{\mathop{\rm tr}}
\newcommand{\half}{{1\over 2}}
\newcommand{\third}{{1\over 3}}
\newcommand{\be}{\begin{equation}}
\newcommand{\ee}{\end{equation}}
\newcommand{\bea}{\begin{eqnarray}}
\newcommand{\eea}{\end{eqnarray}}

\newcommand{\dint}[2]{\int\limits_{#1}^{#2}}
\newcommand{\D}{\displaystyle}
\newcommand{\PDT}[1]{\frac{\partial #1}{\partial t}}
\newcommand{\PD}{\partial}
\newcommand{\tw}{\tilde{w}}
\newcommand{\tg}{\tilde{g}}
\newcommand{\newcaption}[1]{\centerline{\parbox{6in}{\caption{#1}}}}

\section{Introduction}

The defining property of Dirichlet $p$-branes in type II superstring
theory is that fundamental strings can have endpoints inside a
$p$-brane worldvolume, whereas in the bulk of spacetime, away from any
$p$-branes, the theory only contains closed strings.  D-brane dynamics
is governed by worldvolume terms in the spacetime effective action of
type II string theory which arise from the open-string sector of the
worldsheet theory.  In a flat supergravity background the worldvolume
action is given by the dimensional reduction of a ten-dimensional
${\cal{N}}=1$ supersymmetric gauge theory.  In general this is some
complicated non-abelian theory involving high-derivative terms whose
detailed structure is not known at present, but for many applications
it may be approximated by supersymmetric Yang-Mills theory, which is
then dimensionally reduced in a straightforward manner.  For the
special case of a single $p$-brane the worldvolume gauge theory is
abelian and is approximated, for slowly varying fields, by the
dimensional reduction of ten-dimensional supersymmetric Born-Infeld
non-linear electrodynamics.  We also expect the Born-Infeld theory to
govern any dynamics that only involves an abelian subsector of the
more complicated non-abelian theory.

The non-linear Born-Infeld electrodynamics can itself be approximated 
by a linear theory, which is just ordinary Maxwell electrodynamics, but in 
doing so one loses some interesting features.  It was for example recently
shown, by Callan and Maldacena \cite{calmal}, by Gibbons \cite{gibbons},
and by Howe {\it et al.} \cite{howeetal},
that the full non-linear Born-Infeld theory has simple classical solutions
with the interpretation of macroscopic strings extending from the $p$-brane
worldvolume.  It is interesting to see the open fundamental string arise as
a classical solution in a gauge theory which was obtained from open string
dynamics in the first place.

In the present paper we study these strings further in a number of ways.
We first show how the Born-Infeld solution, describing some number of
parallel fundamental strings emanating from a single $p$-brane, is 
generalized to the case of fundamental strings extending from a collection
of parallel $p$-branes.  This configuration lives in an abelian 
subsector of the gauge theory and does not involve non-abelian 
directions, and so does not require knowledge of the full non-abelian 
theory.

We then turn our attention to the supersymmetry of these
configurations.  We explicitly show that the string solutions of the
Born-Infeld theory satisfy the supersymmetry conditions of the full
non-linear theory, with the fraction of unbroken supersymmetry being
$\nu =1/4$.

Finally, we address some dynamical issues.  Callan and Maldacena
\cite{calmal} considered fluctuations in the background of a string
attached to a three-brane.  They obtained a wave equation for the
propagation of modes that are transverse to both the three-brane
worldvolume and the string.  In the bulk of the three-brane worldvolume the
equation describes free propagation in $3+1$ dimensions but along the
string it reduces to the $1+1$-dimensional wave equation.  In the throat
region where the string attaches to the three-brane the transverse modes
encounter a potential which leads to non-trivial reflection and
transmission amplitudes.  In the low-frequency limit, Callan and Maldacena
found that incident modes on the string undergo perfect reflection from the
three-brane in a manner consistent with a Dirichlet boundary condition.

We obtain the corresponding equations for transverse fluctuations in 
the case of a general $p$-brane with string attached.  We then study an
analogous problem in supergravity, with a macroscopic string extending
from an extremal black $p$-brane.  Unfortunately, we do not have the
appropriate solution of the full supergravity equations at our disposal,
so we instead work within a test-string approximation.  In this approximation
we consider a macroscopic fundamental string in the curved background
geometry of an extremal $p$-brane and obtain a wave-equation for 
transverse fluctuations traveling along the string.  Remarkably, for $p=3,4$, 
the two approaches give identical results upon identification of certain
parameters.  We finish with a discussion of the range of validity of the
two calculations and some suggestions for further work.
 
\section{Fundamental strings emanating from D-branes}

The classical dynamics of a Dirichlet $p$-brane are governed by a
non-linear effective action.  Kappa-symmetric gauge-invariant actions for
Dirichlet $p$-branes have been constructed for arbitrary supergravity
backgrounds in \cite{cederetal} and \cite{bergtown}.  In a remarkable way,
the requirement of kappa-symmetry puts the supergravity background
on-shell.  In this section, however, we will be studying D$p$-branes in the
weak-coupling regime where the supergravity background may be taken to be
flat.  Kappa- and super-symmetry were studied for flat backgrounds, and in
detail in the static gauge, by Aganagic {\it et al.}\ \cite{jhsetal}.  For
our purposes the static gauge will be convenient, and so we will use the
results of \cite{jhsetal}.

Our conventions are essentially similar to those of that work, but not
identical, so we list them here.  The spacetime signature is 
$(-,+,\ldots,+)$.  Greek indices are curved, and roman indices are flat.  
Indices $\alpha,\beta, a,b = 0,\ldots p$ are worldbrane;
$\rho,\sigma,r,s=(p+1)\ldots 9$ are transverse to the brane; and
$\mu,\nu,m,n=0\ldots 9$ are ten dimensional.  Where we need to enumerate
indices and confusion could arise, we will use hats to denote flat indices.
Worldbrane coordinates are $\sigma^\alpha$, target space coordinates are
$x^\mu$.  Lastly, we take $\alpha^\prime=1$.

The action for the D$p$-brane then has two terms, the first being the
Born-Infeld term and the second the Wess-Zumino term which allows
branes to end on other branes.  The relative coefficients are fixed by
kappa-symmetry.

The static gauge is an embedding choice which sets the $p+1$ longitudinal
target, and worldbrane, coordinates equal:
\begin{equation}
x^\alpha = \sigma^\alpha \quad .
\end{equation}

Kappa symmetry may be used to eliminate half of the fermionic degrees of
freedom; in static gauge one of the two spinors of the type II theory may
be set to zero.  This has the consequence that the WZ term vanishes in
static gauge.  The remaining spinor is renamed $\lambda$, and is 
the superpartner of the gauge and scalar fields.

\subsection{Strings emanating from a lone D-brane}

The equation of motion for a purely bosonic background in flat spacetime, 
using ten-dimensional notation, is
\begin{equation}
\Bigl( {1\over \eta - F^2} \Bigr)_\lambda^{\ \nu}
\partial_\nu F_\mu^{\ \lambda} = 0 \quad ,
\label{bieq}
\end{equation}
where $F_{\alpha\beta}=\partial_\alpha A_\beta-\partial_\beta A_\alpha$ is
the field strength of an abelian gauge field living in the $p$-brane
worldvolume.  The remaining components of $F_{\mu\nu}$ are
$F_{\alpha\rho}=-\partial_\alpha\phi_\rho$, where
$\phi_\rho(\sigma^\alpha)$ are scalar fields describing the transverse
displacement of the $p$-brane worldvolume, and finally $F_{\rho\sigma}=0$
reflects the fact that all fields in the action depend only on worldvolume
coordinates.

The Born-Infeld equation (\ref{bieq}) has many interesting solutions.  We
focus on a class of static solutions, described recently by Callan and
Maldacena \cite{calmal} and by Gibbons \cite{gibbons}, which have the
interpretation of macroscopic fundamental strings extending away from a
$p$-brane in a perpendicular direction (closely related solutions were 
also obtained by Howe {\it et al.} \cite{howeetal}).  
Such a solution may be obtained by a
simple ansatz, describing the electric field $E_\alpha$ due to a point
charge in the $p$-brane worldvolume with only one of the scalar fields
excited, say $\phi^{9}$, such that $\partial_\alpha \phi^9 = \pm E_\alpha$.
The choice of sign determines whether the string extends in the positive or
negative $x^9$ direction.  This solution is easily extended to include 
strings protruding from different locations on the $p$-brane
\cite{calmal,gibbons}.

With this ansatz one needs to keep track of only  the $\mu,\nu
=0,\ldots ,p,9$ components of the field strength tensor, which we combine
into a $p+2$ dimensional matrix 
\begin{equation} F_\mu^{\ \nu} = \left[
\begin{array}{ccc}
0 & \pm {\vec{\partial}} \phi^9 & 0 \\
\pm {\vec{\partial}} \phi^9 & 0 & -{\vec{\partial}}\phi^9 \\
0 & +{\vec{\partial}} \phi^9 & 0 
\end{array}
\right] \quad ,
\label{backg}
\end{equation}
where $\vec{\partial}$ denotes the spatial gradient on the worldvolume.
The Born-Infeld equation 
(\ref{bieq}) then reduces to 
\begin{equation}
{\vec{\partial}}^{\, 2}\phi^9 =0 \  ,
\end{equation}
which is solved by the $p$-dimensional Coulomb potential
\begin{equation}
\phi^9(r) = \pm {b_p \over p-2} \, {1\over r^{p-2}} \quad ,
\label{coulomb}
\end{equation} 
where $r = \sqrt{\vec{x}^{\, 2}}$ is a radial worldvolume coordinate.  
For $p=2$ the potential is logarithmic so we will only consider $p\geq 3$.

Since $\phi^9$ has the interpretation of the location of the brane in the
$\hat 9$ direction, we see that the D$p$-brane has developed a
one-dimensional spike, emanating from the origin of worldbrane coordinates
$r=0$, directed along $x^9$.  The spike is interpreted as a string
\cite{calmal,gibbons}.  This identification was solidified in \cite{calmal}
where it was shown that $b_p$ is quantized such that the energy per
unit length of the spike is the fundamental string tension.  The solution
(\ref{coulomb}) is generalized to the case of multiple strings extending
from $r=0$ by multiplying the unit of charge $b_p$ by the number of
fundamental strings $N_1$.

\subsection{Multiple D-branes}

The discussion so far has been limited to the case where the number of
$p$-branes was $N_p=1$.  It is natural to ask whether or not this 
construction can be carried over to the case where there are 
multiple $p$-branes.  Later on we will compare gauge theory results 
to classical supergravity calculations which are only reliable if the 
$p$-brane carries a macroscopic R-R charge, {\it i.e.} it is a collection
of many coincident parallel branes.

For $N_{p} > 1$, the worldvolume theory is described by a
non-abelian $U(N_{p})$ theory, in which the transverse coordinates 
$\phi^i$ become matrix-valued.  In abelian subsectors this action is 
expected to be just the Born-Infeld action again.  Here we will look for 
the ground state of the multi-string-brane system for given $N_1$ and $N_{p}$.

In our solutions, $\partial_r\phi^9=\pm F_{0r}$, so the $U(N_p)$ matrices
$F_{0r}$ and $\phi^9$ may be simultaneously diagonalized, and since we
do not excite any other directions we do not need to worry about subtleties
of the non-abelian generalization of the Born-Infeld action.  
Each diagonal entry of $\phi^9$ then represents
the $x^9$ coordinate of each $p$-brane.  For now, let $N_1$ be larger than,
and divisible by, $N_p$, and imagine distributing the fundamental strings
among the $p$-branes.  Put $n_i$ strings on the i-th $p$-brane such that
\begin{equation}
N_1 = \sum\limits_{i=1}^{N_p} n_i \quad .  
\end{equation}
Then the diagonal entries of $\phi^9$ are
\be
\phi^9_i(r) = \frac{n_i}{p-2}\frac{b_p}{r^{p-2}} \quad .
\ee
Now let us introduce a cutoff $\delta$ on the worldbrane radial coordinate
$r$ to keep the string length finite, and calculate the total energy of the
system,
\be
E(\delta) = 
\D\left\{ N_1 + {N_p^2\over N_1}(\Delta n)^2 \right\} T_f \phi^9_{cm}(\delta)
\quad .
\label{totalenergy}
\ee 
Here $T_f$ is the fundamental string tension, 
$\phi^9_{cm}(r)={1\over N_p} \tr \phi^9(r)$ is the center of mass coordinate 
of the strings, and $\Delta n$ is the standard deviation of the 
distribution $\{n_i\}$.
It follows that the energy is minimized when the strings are evenly 
distributed among the branes. The minimum energy is proportional to the 
center of mass coordinate of the strings. 

In this naive picture, the exact 
ground state energy can be obtained only when $N_1$ is divisible by 
$N_p$.  It is, however, possible for the branes to ``share'' the strings for 
arbitrary $N_1$ in the following way.  
In the gauge theory, separate the center-of-mass $U(1)$ from $U(N_p)
= [U(1)\times SU(N_p)]/Z_n$. The energy density due to the 
center-of-mass electric field is $(N_p/2g_p)F^2$, where the factor 
of $N_p$ can be understood as the total mass of the mechanical analogue. 
To obtain the minimum value of the energy, we recall that flux quantization 
is modified in the presence of multiple coincident $p$-branes.  
Fractionation occurs and the unit electric flux on each brane 
is reduced by $1/N_p$ by comparison with the case of the lone brane.

Then, since the unit charge for the single brane is $b_p$, the center of
mass transverse position and electric field become 
\be 
\phi^9_{cm} (r) = \pm {1\over p-2}{N_1\over N_p} {b_p\over r^{p-2}}\ ,
\quad
F_{0r}^{cm} = {N_1\over N_p}{b_p\over r^{p-2}} \ .
\label{comcom}
\ee
The ground state energy of the system is
\be
E(\delta) = N_1 T_f \phi^9_{cm}(\delta) \quad ,
\label{comenergy}
\ee
which exactly equals the ground state energy in (\ref{totalenergy}), but 
is no longer subject to the same restrictions on $N_1$ and $N_p$.

The upshot of all this is that whenever we consider a multiple-brane spike, 
we can work entirely within the center-of-mass $U(1)$ and the equations 
are identical to those for the lone-brane spike.  We only have to remember 
to replace the unit charge $b_p$ by $b_p (N_1/N_p)$.

\section{Supersymmetry}

In \cite{calmal,gibbons} it was conjectured on the basis of calculations in
the Maxwell limit that the brane-spike configuration is supersymmetric.  We
now proceed to show how this works in the full Born-Infeld theory.  We will
work explicitly with the case $N_p=1$ but the calculation would be
identical for the multiple-brane case by the reasoning at the end of the
last section.

For the purpose of investigating supersymmetry, it is convenient to use the
dimensionally reduced form of the Born-Infeld action.  Again we follow
\cite{jhsetal}.  The action is
\begin{equation}
S = -T_{p} \int d^{p+1}\sigma
\sqrt{-\det\left(G_{\alpha\beta}^{(p)}
+{\cal{F}}_{\alpha\beta}^{(p)}\right)} \quad ,
\end{equation}
where 
\begin{eqnarray}  
G_{\alpha\beta}^{(p)} &=& \eta_{\alpha\beta} + \partial_\alpha\phi_\rho
\partial_\beta\phi^\rho - {\bar{\lambda}}
(\Gamma_\alpha+\Gamma_\rho\partial_\alpha\phi^\rho) \partial_\beta\lambda 
\nonumber \\
& & -{\bar{\lambda}} (\Gamma_\beta+\Gamma_\rho\partial_\beta\phi^\rho)
\partial_\alpha\lambda + {\bar{\lambda}}\Gamma^\mu\partial_\alpha\lambda
{\bar{\lambda}}\Gamma_\mu\partial_\beta\lambda  \quad , \\
\nonumber\\
{\cal{F}}_{\alpha\beta}^{(p)} &=& F_{\alpha\beta} -{\bar{\lambda}}
(\Gamma_\alpha+\Gamma_\rho\partial_\alpha\phi^\rho) \partial_\beta\lambda
+{\bar{\lambda}} (\Gamma_\alpha+\Gamma_\rho\partial_\alpha\phi^\rho)
\partial_\beta\lambda  \quad .
\end{eqnarray}
In these expressions, $\lambda$ are the the fermionic superpartners of the
gauge and scalar fields.

The choice of static gauge is consistent with
kappa-symmetry and supersymmetry.  The resulting supersymmetry
variations are \cite{jhsetal}
\begin{eqnarray}
\delta {\bar{\lambda}} &=& {\bar{\epsilon}} \Bigl[ \Delta^{(p)} +
\zeta^{(p)} \Bigr] + \xi^\alpha \partial_\alpha {\bar{\lambda}} \quad ,   \\
\nonumber \\
\delta \phi^\rho &=& {\bar{\epsilon}} \Bigl[ \Delta^{(p)} - \zeta^{(p)} \Bigr]
\Gamma^\rho\lambda + \xi^\alpha\partial_\alpha\phi^\rho \quad ,  \\
\nonumber \\
\delta A_\alpha &=& {\bar{\epsilon}} \Bigl[ \zeta^{(p)} - \Delta^{(p)}
\Bigr] (\Gamma_\alpha+\Gamma_\rho\partial_\alpha\phi^\rho)\lambda +
{\bar{\epsilon}} \Biggl[ \third\Delta^{(p)} - \zeta^{(p)} \Biggr] 
\Gamma_\mu\lambda {\bar{\lambda}}\Gamma^\mu\partial_\alpha\lambda \\
& & + \xi^\rho\partial_\rho{A_\alpha} + \partial_\alpha\xi^\rho A_\rho \quad , 
\end{eqnarray}
where the compensating general coordinate transformation parameter
is \cite{jhsetal}
\begin{equation}
\xi^\alpha={\bar{\epsilon}}(\zeta^{(p)}
-\Delta^{(p)})\Gamma^\alpha\lambda \quad .
\end{equation}
In the above we  have used the abbreviation
\begin{equation}
\Delta^{(p)}\equiv \pm \Gamma_9\Gamma_8\ldots \Gamma_{p+1} \quad ,
\end{equation}
where the $\pm$ sign is determined by reducing from ten dimensions.  The
crucial quantity $\zeta^{(p)}$ depends on the gauge field via $\cal{F}$.

For bosonic backgrounds, the only nontrivial supersymmetry 
transformation is
\begin{equation}
\delta {\bar{\lambda}} = {\bar{\epsilon}} \Bigl[ \Delta^{(p)} +
\zeta^{(p)} \Bigr] \quad .
\end{equation}
In order to proceed further, we need the form of $\zeta^{(p)}$.  We can
extract it from the formul\ae\, in section 3.2 of \cite{jhsetal}.  The
equations are slightly different for the IIA and IIB cases, in that
$\Gamma_{11}$ appears in IIA equations while $SO(2)$ matrices $\{\tau_1,
\tau_3, i\tau_2 \}$ appear in those of IIB.

For definiteness, let us exhibit the relevant equations for
the IIB theory with $p$ odd:
\begin{eqnarray}
\gamma^{(p)} &=& \pmatrix{ 0 & \zeta^{(p)} \cr {\tilde{\zeta}}^{(p)} & 0
\cr} \quad , \\
\rho^{(p)} &=& \sqrt{-\det\left(G+{\cal{F}}\right)} \, \gamma^{(p)} \quad ,\\
\rho^{(p)} &=& {\frac{1}{(p+1)!}}\epsilon^{\alpha_1\ldots\alpha_{p+1}}
\rho_{\alpha_1\ldots\alpha_{p+1}} \quad ,\\
\rho_{p+1} &=& {\frac{1}{(p+1)!}} \rho_{\alpha_1\ldots\alpha_{p+1}} 
d\sigma^{\alpha_1}\ldots d\sigma^{p+1} \quad ,\\
\rho_{B} &=& \sum_{p \, odd}\rho_{p+1} = 
e^{\cal{F}} C_B(\psi) \tau_1 \quad ,\\
C_B(\psi) &=& (\tau_3) + {\frac{1}{2!}}\psi^2 + 
{\frac{1}{4!}}(\tau_3)\psi^4 + {\frac{1}{6!}}\psi^6 + \ldots \quad ,\\
\psi &=& \gamma_\alpha d\sigma^\alpha = (\partial_\alpha X^\mu - 
{\bar{\lambda}}\Gamma^\mu\partial_\alpha\lambda)
\Gamma_\mu d\sigma^\alpha \quad .
\end{eqnarray}
In order to untangle this, let us inspect our background configuration.  We
have that the gauge field is purely electric and that the scalar $\phi^9$
is related to it by
\begin{equation}
\partial_r\phi^9 = \chi {F_{0r}} \quad ,
\end{equation}
where $\chi=\pm 1$ and $r$ is the radial variable in the worldbrane.  
For a configuration of this form, the Born-Infeld determinant conveniently
collapses to unity.

In addition, since $F$ is purely electric and our background is bosonic,
the matrices $\rho$ take the form 
\begin{equation}
\rho_{p+1} = {\frac{1}{(p+1)!}} (\tau_3^{(p-1)/2}\tau_1) \psi^{p+1} 
 + {\frac{1}{(p-1)!}} (\tau_3^{(p+1)/2}\tau_1) {\cal{F}} \psi^{p-1}
\quad .
\end{equation}

{}From now on, we will concentrate on the threebrane for definiteness, but
the supersymmetry equations for other $p=4,\ldots,9$ follow in an exactly
analogous fashion.  Of course, this is to be expected from T-duality.  

For the threebrane,
\begin{equation}
\rho_4 = {\frac{1}{4!}}(\tau_3\tau_1)\psi^4 
+ {\frac{1}{2!}}(\tau_1){\cal{F}}\psi^2 \quad ,
\end{equation}
and so $\zeta^{(3)}$ becomes
\begin{equation}
\zeta^{(3)}  = \epsilon^{\alpha_1\alpha_2\alpha_3\alpha_4} \Biggl[
{\frac{1}{4!}}
\gamma_{\alpha_1}\gamma_{\alpha_2}\gamma_{\alpha_3}\gamma_{\alpha_4} +
{\frac{1}{2!}}
{\cal{F}}_{\alpha_1 \alpha_2}\gamma_{\alpha_3}\gamma_{\alpha_4}
\Biggr] \quad ,
\end{equation}
where
\begin{eqnarray}
\gamma_\alpha &=& \left(\partial_\alpha X^\mu -
{\bar{\lambda}}\Gamma^\mu\partial_\alpha\lambda\right) \Gamma_\mu \\
&=& \Gamma_\alpha+\Gamma_\rho\partial_\alpha\phi^\rho \quad ,
\end{eqnarray}
because we are in a bosonic background and in static gauge.  Recall also
that in a bosonic background ${\cal{F}} = F$.

Our supersymmetry equation is now

\begin{equation}
\delta{\bar{\lambda}} = {\bar{\epsilon}} \Biggl[ F_{{\hat{0}}{\hat{r}}} 
\Bigl(1+\chi(\Gamma_9\Gamma_0)\Bigr) 
\Gamma_{\hat{\theta}} \Gamma_{\hat{\phi}} \Biggr] \quad .
\end{equation}
To get this, we used the fact that the IIB spinor $\epsilon$ obeys
the chirality condition $\Gamma_{11}\epsilon=\epsilon$.

Let the spinor $\epsilon$ also obey the condition
\begin{equation}\label{f1susy} 
(\Gamma_0\Gamma_9)\epsilon = \chi_1 \epsilon 
\end{equation} 
appropriate to a string, with $\chi_{1} = \pm 1$.

Then we have unbroken supersymmetry for
\begin{equation}
\chi = -\chi_1 \quad .
\end{equation}

We see from (\ref{f1susy}) that the fraction of supersymmetry unbroken
by the brane with spike is $\nu=1/4$, as expected.

\section{Brane-waves from Born-Infeld}

In this section we will consider perturbations in the background of a
$p$-brane with a string attached, from the worldbrane point of view.
\footnote{Scattering of waves traveling along a string attached to a D-brane has been considered from a different point of view by 
Giddings\cite{giddings}.}
Let us expand the ten dimensional gauge field as
\be
F_\mu^{\ \nu} = \bar{F}_\mu^{\ \nu} + \delta F_\mu^{\ \nu} \quad ,
\ee
where $\bar{F}_\mu^{\ \nu}$ is the background value (\ref{backg}) and
the fluctuations can in general have components both parallel and
transverse to the $p$-brane,
\begin{eqnarray}
\delta F_{\alpha\beta} &=& \partial_\alpha (\delta A_\beta) -
\partial_\beta (\delta A_\alpha ) \quad , \\ 
\delta F_{\alpha\rho} &=&
-\partial_\alpha (\delta \phi_\rho) \quad .
\end{eqnarray}
For the multiple brane case we work within the center of mass $U(1)$.
Let $B_\mu^{\ \nu}$ be the matrix prefactor to the derivative of $F$ in the
Born-Infeld equation of motion (\ref{bieq}).  
Then to first order in fluctuations,
\bea
0 &=& \bar{B}_\lambda^{\ \nu} \partial_\nu (\delta F)_\mu^{\ \lambda}
+ (\delta B)_\lambda^{\ \nu} \partial_\nu \bar{F}_\mu^{\ \lambda}  
\nonumber \\
&=& \bar{B}_\lambda^{\ \nu} \partial_\nu (\delta F)_\mu^{\ \lambda}
+ (\bar{B}[\bar{F}\,\delta F + \delta F\, \bar{F}]\bar{B})_\lambda^{\ \nu}
\partial_\nu \bar{F}_\mu^{\ \lambda} \quad .
\label{fluct}
\eea

The simplest case to consider is perturbations $\delta\phi^\perp(x^\alpha)$,
that are transverse to both the string and the $p$-brane.  The fluctuation
equation then collapses to
\be
\left[ \left(1+(\vec{\partial}\phi^9)^2\right) \partial_t^2
-\vec{\partial}^{\,2} \right] \delta\phi^\perp = 0 \quad .
\label{fluctii}
\ee
This equation has several interesting aspects.  The first, as noted for
three-branes in \cite{calmal}, is that ripples in transverse position 
satisfy the appropriate one-dimensional wave equation along the string.  
To see this, let us switch the radial coordinate to
\begin{equation}
u = {\frac{b_p (N_1/N_p)}{(p-2)r^{p-2}}} \quad ,
\end{equation}
for which the background scalar takes the simple form $\phi^9_{cm}=u$.  The
region near $u\rightarrow\infty$ ({\it i.e.} $r\rightarrow 0$) describes
the string-like spike and there the fluctuation equation reduces to
\begin{equation}
\Bigl( \partial_t^2 - \partial_u^2 \Bigr) \delta\phi^\perp = 0 \quad .
\end{equation}
The worldvolume angular information is irrelevant in the
$u\rightarrow\infty$ region, and the equation becomes
two-dimensional.  This further supports the identification of the 
the brane spike as a string in this region.

In the region $u\rightarrow 0$ ({\it i.e.} $r\rightarrow \infty$), on
the other hand, the fluctuation equation (\ref{fluctii}) becomes the
free wave equation in $p$ spatial dimensions and describes wave
propagation in the bulk of the $p$-brane worldvolume.  The most
interesting region is in between where the transition between bulk
$p$-brane behavior and one-dimensional string behavior occurs.

In the following section, we will compare these transverse fluctuations
in Born-Infeld theory to a supergravity calculation of waves propagating
along a string attached to a $p$-brane.  The level of approximation in
the gravity calculation is to treat the string as a test string in the
curved background geometry of a $p$-brane.  The test string can 
carry no information about the angular variables of the worldvolume
and accordingly we will restrict to s-waves in the remainder of this
section.

Anticipating the supergravity result, we will also restrict the value
of $p$.  The three-brane is known to have a core region which is 
nonsingular and so we expect that the worldbrane and supergravity 
perturbation equations have a chance of agreeing.  Similarly, in the case 
of the four-brane we expect that agreement is possible\footnote{We thank 
Igor Klebanov for a discussion on this point.}. 
The reasoning is that the fourbrane-string system
may be thought of as the dimensional reduction of the M-theory five-brane
with an M-theory two-brane attached and the core region of the five-brane
geometry is also known to be nonsingular.  Thus we will concentrate on 
the cases $p=3,4$ from now on.

A mode of frequency $\omega$ satisfies
\be
\left[{d^2\over dy^2} + 1 + a_p y^{-{2p-2\over p-2}}\right]
\delta\phi^\perp =0  \quad ,
\ee
where 
\be
a_p = \bigl({\omega\over p-2}\bigr)^{2p-2\over p-2} \ 
b_p^{2\over p-2} \quad ,
\ee
and we have defined a rescaled variable $y=\omega u$.  Note that the 
power of $y$ that appears in the differential equation is integer valued 
for $p=3,4$ only, which is another indication that these values are
special.

At this point it is useful to define a new coordinate, 
$\xi \in  (-\infty,+\infty)$, which blows up the $y\sim 0$ region.
Our choice (for $p=3,4$) is
\begin{equation}
{\frac{d\xi}{dy}} = \sqrt{h_p(y)} \quad ,
\label{bixi}
\end{equation}
where 
\begin{equation}
h_p(y) = \Bigl(1 + {a_p\over y^{7-p}} \Bigr) \quad .
\end{equation}
Due to the change of variables from $y$ to $\xi$, our differential
equation develops a term with a linear $\xi$ derivative.  In order to 
get rid of it, we rescale $\delta\phi^\perp$ as follows,
\begin{equation}
\delta\phi^\perp_{(p)} = h_p(y)^{-1/4} \, 
\delta{\tilde{\phi}}^\perp_{(p)} \quad .
\label{wkbrescaling}
\end{equation}
The equation for the transverse brane fluctuations takes the form of 
a one-dimensional Schr\"odinger equation,
\begin{equation}
\Biggl[ -{\frac{d^2}{d\xi^2}}  + V_{(p)} \Biggr]
\delta{\tilde{\phi}}^\perp_{(p)} = \delta{\tilde{\phi}}^\perp_{(p)}  \quad ,
\end{equation}
with the potential given by
\begin{eqnarray}
V_{(3)}(\xi) &=& {\frac{5a_3}{y^6}} h_3(y)^{-3}
\label{wbv3} \quad , \\ 
\nonumber\\
V_{(4)}(\xi) &=&  {\frac{3a_4}{y^5}} 
\Biggl[ 1 + {\frac{a_4}{16 y^3}} \Biggr]h_4(y)^{-3} \quad , 
\label{wbv4}
\end{eqnarray}
where $a_3 = \omega^4 (b_3 N_1/N_3)^2$ and $a_4 = \omega^3 (b_4
N_1/N_4)/8$.

Callan and Maldacena \cite{calmal} analyzed the scattering problem for 
the three-brane by approximating the potential by a delta function in the 
long wavelength limit. They found that the reflection amplitude $R=-1$ in 
that limit and interpreted it as a  Dirichlet boundary condition on string
modes incident on the three-brane.  
The same argument goes through for the four-brane case, although the 
details of the calculation are somewhat different.

\section{Test string in a $p$-brane supergravity background}

The full supergravity solution for a string extending from a $p$-brane at
right angles is not known.  We and others
\cite{JPG} have attempted to construct such solutions, but the closest we
have got is to U-dualize a solution of \cite{AAT} in which one or other of
the branes is partially delocalized.   This solution reduces to a pure
$p$-brane geometry or a pure string geometry in appropriate limits
but does not capture the physics of the two together in a suitable
fashion for our present discussion.   In the absence of a fully localized
supergravity solution we will make do with a test-string approximation.
We will treat the $p$-brane as a supergravity background for a 
calculation of wave propagation along a macroscopic fundamental
string, which is embedded in the $p$-brane geometry.  An analogous
treatment of a string extending out from a Schwarzschild black hole
has been carried out by Lawrence and Martinec \cite{lawmar}.

Let the string lie along a radial direction $u$ in the space transverse
to the $p$-brane, and use static gauge:
\begin{eqnarray}
X_{cl}^0(\tau,\sigma) &=& \tau \quad ,\\
X_{cl}^u(\tau,\sigma) &=& \sigma \quad ,\\
X_{cl}^\parallel(\tau,\sigma) &=& 0 \quad ,\\
X_{cl}^\perp(\tau,\sigma) &=& 0 \quad .
\end{eqnarray}
Here $X^\parallel$ denotes the coordinates longitudinal to the threebrane
but perpendicular to the string, and $X^\perp$ denotes the purely 
transverse coordinates.   

The motion of a string in a supergravity background is governed by the
usual two-dimensional sigma-model Lagrangian.  
In studying fluctuations of a string around
a classical position as given above, it is most convenient to use the
normal coordinate expansion and work in tangent space rather than curved
space.  The relevant part of the sigma-model action
to second order in perturbations is
\cite{mukhietal}
\be
S = {\frac{1}{2}} \int d^2\sigma \sqrt{-h} \left[ \eta_{m n}
h^{\alpha\beta} (D_\alpha \eta)^m (D_\beta \eta)^n 
+R_{\mu m n \nu} \eta^m \eta^n h^{\alpha\beta} \partial_\alpha
X_{cl}^\mu \partial_\beta X_{cl}^\nu \right]  \quad .
\ee
In this action, $h_{\alpha\beta}$ is the intrinsic worldsheet metric, which
in static gauge becomes the pullback of the background metric to the
worldsheet.  The variables $\eta^m$ are the tangent space 
Riemann normal coordinates,
\begin{equation}
\eta^m = e_\mu^m(X_{cl}) \eta^\mu \quad ,
\end{equation}
whose covariant derivatives in tangent space are defined via the
spin-connection $\omega_\mu^{[m n]}$ as
\begin{equation}
(D_\alpha \eta)^m = \partial_\alpha \eta^m + \omega_\mu^{m
n}\partial_\alpha X^\mu \eta_n \quad .
\end{equation}
We ignore the ghost sector because we are only interested in
transverse fluctuations $\eta^\perp$.  We also ignore the dilaton term 
in the worldsheet action because
it turns out not to contibute to transverse fluctuation
equations.  There is no antisymmetric tensor term present in the
action because we are working in a $p$-brane 
background with only Ramond-Ramond charge.

The next step toward finding the equations for transverse fluctuations is
to calculate the various components of the spin-connection and the 
Riemann tensor in the $p$-brane supergravity background.  
For this, we need the metric.  In string frame, it is
\begin{equation}
ds^2 = H_p(u)^{-1/2} [-dt^2 + (d{{\vec{x}}^\parallel})^2 ] + H_p(u)^{1/2}
 (d{{\vec{x}}^\perp})^2 \quad ,
\end{equation}
where
\begin{equation}
H_p(u) = 1 + {\frac{R_p^{7-p}}{u^{7-p}}} \quad ,
\end{equation}
$R_p$ is essentially the gravitational radius of the $p$-brane,
and $u=|{\vec{x}^\perp}|$ is the radial coordinate of the transverse
space.  
There is also a dilaton present, $e^{-2\Phi}=H_p(u)^{(p-3)/2}$, which is
constant in the case of a three-brane. 

Before we calculate the spin-connections and curvatures, we change to
tortoise-type coordinates via
\begin{equation}
{\frac{d\xi}{du}} = \sqrt{H_p(u)} \quad ,
\label{sgxi}
\end{equation}
so that the coordinate along the string runs over the whole 
real axis, $\xi \in (-\infty,\infty)$, and the metric takes the form,
\begin{equation}
ds^2 = 
H_p(u)^{-1/2} [-dt^2 + (d{{\vec{x}}^\parallel})^2 + d\xi^2] + 
H_p(u)^{1/2} \,u^2 \, d\Omega^2_{8-p} \quad .
\end{equation}
Note the similarity between (\ref{sgxi}) and the analogous redefinition
(\ref{bixi}) in the gauge theory.

We then find that the relevant components of the spin-connection are
\begin{eqnarray}
\omega_{\hat{0}}^{\ \hat{0} \, \hat{\xi}} &=
&{\frac{(7-p)}{4}}{\frac{R_p^{7-p}}{u^{8-p}}} H_p(u)^{-5/4} \quad , \\
\omega_{\hat{r}}^{\ \hat{s} \, \hat{\xi}} 
&=& \delta_{\hat{r}}^{\hat{s}} {\frac{1}{u}}
\left[1+{\frac{(p-3)}{4}}{\frac{R_p^{7-p}}{u^{7-p}}} \right] 
H_p(u)^{-5/4} \quad ,
\end{eqnarray}
where $\hat{r},\hat{s}$ denote directions transverse to both the string and
$p$-brane.  The components of the Riemann tensor needed for our calculation
are
\begin{eqnarray}
R^{(p)}_{t \, \hat{r} \, \hat{s} \, t} &=& -\delta_{\hat{r}\hat{s}}{\frac{(7-p)}{4}} 
{\frac{R_p^{7-p}}{u^{9-p}}} \Biggl[ 1 + {\frac{(p-3)}{4}}
{\frac{R_p^{7-p}}{u^{7-p}}} \Biggr] H_p(u)^{-3} \quad , \\ 
R^{(p)}_{\xi \, \hat{r} \, \hat{s} \, \xi} &=& 
+\delta_{\hat{r}\hat{s}}{\frac{(7-p)^2}{4}}
 {\frac{R_p^{7-p}}{u^{9-p}}} H_p(u)^{-3} \quad .
\end{eqnarray}
The resulting equation for a transverse fluctuation mode of 
frequency $\omega$, using rescaled variables $\tilde{\xi}=\omega\xi$
and $y=\omega u$, takes the form
\begin{equation}
\Biggl[ - {d^2\over d\tilde{\xi}^2} + V_{(p)} \Biggl]
\eta^{\hat{s}}_{(p)}   = \eta^{\hat{s}}_{(p)} \quad ,
\end{equation}
where
\begin{equation}
V_{(p)}(\tilde{\xi}) = {\frac{1}{4}} (7-p)(8-p) 
{\frac{(\omega R_p)^{7-p}}{y^{9-p}}} \Biggl[ 1 +
{\frac{(p-3)}{4(8-p)}}
{\frac{(\omega R_p)^{7-p}}{y^{7-p}}} \Biggr]H_p(y)^{-3} \quad .
\end{equation}
For the cases $p=3,4$, in particular, the potentials are
\begin{eqnarray}
V_{(3)}(\tilde{\xi}) &=& {\frac{5(\omega R_3)^4}{y^6}}H_3(y)^{-3} \quad , \\ 
V_{(4)}(\tilde{\xi}) &=& {\frac{3(\omega R_4)^3}{y^5}} \Biggl[ 1 +
{\frac{(\omega R_4)^3}{16y^3}} \Biggr]H_4(y)^{-3} \quad ,
\end{eqnarray}
which is identical to the gauge theory results (\ref{wbv3}) and
(\ref{wbv4}), provided we make the frequency-independent
identifications
\be
R_3^4 \leftrightarrow  [b_3 (N_1/N_3)]^2 \quad ,
\qquad
R_4^3 \leftrightarrow  [b_4 (N_1/N_4)]/8 \quad .
\label{ids}
\ee
In addition to giving the same end result, the gauge theory parallels
the supergravity calculation in a number of ways.  For example, the
scale factor multiplying the gauge theory fluctuation in 
(\ref{wkbrescaling}) corresponds precisely to the appropriate 
component of the zehnbein that relates tangent space and curved
space fluctuations of the test string.

It is remarkable that the Born-Infeld theory, a gauge theory 
formulated in flat spacetime, is able to correctly describe dynamics
that arise from the curved geometry of a $p$-brane on the 
supergravity side.  It is equally remarkable that fluctuations on a
test string deep in the throat of the $p$-brane geometry correspond,
in the gauge theory, to bulk wave propagation in the $p$-brane
worldvolume.

\section{Discussion}

An obvious question is whether the identification of parameters (\ref{ids})
is physically sensible.  In spite of the exact agreement of the form of the
fluctuation equations, the actual values of the parameters have markedly
different physical orgins in the two approaches.

The quantity $b_p$ controls the strength of the electric field on the
brane, and it is directly related to the tension of a fundamental string
\cite{calmal}.  In our conventions,
\begin{equation}
b_p = {\frac{(2\pi)^{p-1}}{\Omega_{p-1}}} g \quad .
\end{equation}
On the other hand, the gravitational quantities $R_3^4,R_4^3$ in
the test string approach are related to the gravitational size of the
supergravity $p$-brane.  In order to get the precise coefficient, it is
simplest to compare the known ADM mass of the supergravity $p$-brane to the
known tension of a D$p$-brane.  For the first quantity, we have \cite{jxlu}
\begin{equation}
M_{ADM} = {\frac{\Omega_{8-p}}{2\kappa_{10}^2}}(7-p)R_p^{7-p}{\mbox{Vol}}_p
\quad ,
\end{equation}
where Vol$_p$ is the volume of the brane, whereas for the second we have
\cite{joetasi}
\begin{equation}
{\frac{M}{{\mbox{Vol}}_{p}}} = {\frac{\sqrt{\pi}}{\kappa_{10}}}
{\frac{1}{(2\pi)^{p-3}}} N_p \quad .
\end{equation}
Therefore, the supergravity parameter $R_p$ is related to the number of
$p$-branes $N_p$ by:
\begin{equation}
R_p^{7-p} = {\frac{2\sqrt{\pi}}{\Omega_{8-p}(2\pi)^{p-3}(7-p)}} 
\kappa_{10} N_p \quad .
\end{equation}
Note also the relationship between the gravitational radius of the
fundamental string, $R_{F1}$, and the number of strings:
\begin{equation}
R_{F1}^6 = {\frac{1}{2\pi^5}} \kappa_{10}^2 N_1 \quad .
\end{equation}
Putting all of this together, and using the fact that $\alpha^\prime=1$
implies that $2\kappa_{10}^2=(2\pi)^7g^2$, we find
\begin{eqnarray}
N_3^3 &=& g \pi^3 N_1^2 \quad ,\\
N_4^2 &=& \pi N_1 \quad .
\end{eqnarray}
The physical meaning of this identification is not entirely transparent.
We may get a handle on it by rewriting it, up to pure numbers of order one,
as
\begin{eqnarray}
(g^2 N_1)^2 &\sim& (g N_3)^3 \quad , \label{puzz3}\\
(g^2 N_1)   &\sim& (g N_4)^2  \quad .
\label{puzz4}
\end{eqnarray}
In our worldvolume approach we were assuming that the supergravity
background was flat.  In order for this to be a good approximation, we need
for the gravitational radii of the D-branes to be less than the
string scale.  In addition, for our identification of the brane spike to be
consistent, we need that the gravitational radius of the fundamental
string be less than the string scale.  In our unit conventions, these
conditions read
\begin{equation}\label{puzzin1}
g N_3 \ll 1 \quad ,  \qquad g N_4 \ll 1 \quad , \qquad g^2 N_1 \ll 1 \quad .
\end{equation}
On the supergravity side, we treated the string as a test string in the
background of the much more massive three- and four-branes.  This is a good
approximation if the gravitational influence of the string is much smaller
than that of the supergravity $p$-brane:
\begin{equation}\label{puzzin2}
g^2 N_1 \ll g N_3 \quad , \qquad g^2 N_1 \ll g N_4 \quad .
\end{equation}
{}From (\ref{puzzin1}) and (\ref{puzzin2}), we see that the agreement
conditions (\ref{puzz3}) and (\ref{puzz4}) are satisfied if the gravitational
fields of both the D-brane and the fundamental string are small in string
units.  This is actually a regime in which we cannot trust the
supergravity solution, but it is intriguing that naive extrapolation of the
test-string-supergravity result to substringy scales gives agreement with
the Born-Infeld calculation.  Perhaps the fact that the threebrane, and the
M-theory analog of the fourbrane, are nonsingular in their cores is playing
an important role here.

Another reason why the agreement is perhaps unexpected is that the 
brane-spike solutions of Born-Infeld theory are suspect near $r=0$, where
derivatives of the field strength are inevitably large.
It is conceivable that the supersymmetry of these configurations leads them 
to also satisfy the higher-derivative equations of motion, and this would 
explain why the gauge theory captures the correct physics far out along
the string.

In any case, the agreement that we have found suggests further avenues of
investigation.  It would, for example, be interesting to investigate
fermionic fluctuations, for which supersymmetry may play a useful role.  
Another issue to tackle is Hawking radiation along the string subsequent
to adding energy to the BPS system.  We
would also like to find the exact supergravity solution corresponding to
the fundamental string emanating from a D$p$-brane, at least for the BPS
case and with luck for the nonextremal case as well.  Once that is known,
we can allow the string to flex its gravitational muscles.  
This supergravity solution would also be U-dual to the
supergravity solution for a ``T''-shaped M-theory fivebrane, which is 
of interest in its own right.

\section*{Acknowledgements}

We are grateful to C. Callan, I. Klebanov, and W. Taylor for useful 
discussions. L.T. thanks the Science Institute of the University of Iceland 
for hospitality during early stages of this work. A.W.P. wishes to thank the 
Aspen Center for Physics for hospitality. 
The work of S.L. and L.T. was supported in part by a US Department of
Energy Outstanding Junior Investigator Award, DE-FG02-91ER40671.  
The work of A.W.P. was supported in part by NSF grants PHY96-00258 
and PHY-94-07194.

\vskip1cm

\noindent
{\bf Note Added:} \hfill\break
\noindent
After completion of this paper, we were informed by one of the
authors of \cite{howeetal} that supersymmetry of a related class of
Born-Infeld solutions was proven there.


\end{document}